
 \magnification=1200
 \tolerance=6000
 \font\fig cmr8

 \font\fig cmr8
 \def\figure#1#2{\centerline{\bf Fig. #1:}\par\noindent{\fig #2}}
 \catcode`\"=\active\let"=\"
 \def\Dlnp{\delta\!\ln p}
 \def\Dlnpt{\delta\!\ln p_t}
 \def\Dlnx{\delta\!\ln x}
 \def\Dlnt{\delta\!\ln \theta}
 \def\Dlnth{\delta\!\ln \theta_{\rm th}}
 \def\Dlnps{\delta\!\ln \psi}
 \def\De{\delta\! e}
 \def\DE{\delta\! E}
 \def\Df{\Delta\! f}
 \def\Dg{\Delta\! g}
 \def\Dq{\delta\! q}
 \def\ddp#1{{d #1 \over d \phi }}
 \def\ddpp#1{{d #1 \over d \phip }}
 \def\ddpd#1{{d^2 #1 \over d\phi^2}}
 \def\ppp{(\phi ,\phip )}
 \def\p{(\phi )}
 \def\pp{(\phip )}
 \def\zwmA{2\!-\!A}
 \def\drmA{3\!-\!A}
 \def\vimA{4\!-\!A}
 \def\semA{6\!-\!A}
 \def\laf{\lambda_1}
 \def\lag{\lambda_2}
 \def\eps{\varepsilon}
 \def\tth{\theta_{\rm th}}
 \def\xp#1#2{\Bigl(#1-{1\over 2}#2\Bigr)}
 \def\xpp#1#2{\Biggl\{\Bigl(#1\pp-{1\over 2}#2\pp\Bigr)}
 \def\unop{\bigl(1+{1\over 3}\phi\ddp{} \bigr)}
 \def\unopp{\bigl(1+{1\over 3}\phip\ddpp{} \bigr)}
 \def\xpfull#1#2{\biggl\{\Bigl(#1-{1\over 2}#2\Bigr)\cdot\unop
 -{1\over 6}#2\phi\ddp{}\biggr\}}
 \def\xppfull#1#2{\biggl\{\Bigl(#1\pp-{1\over 2}#2\pp\Bigr)
     \cdot\unopp -{1\over 6}#2\pp\phip\ddpp{}\biggr\}}
 \def\LSA#1#2{\Bigl({2\over 3}\phi\ddp{}#1-{4\over 3}#1+#2 \Bigr)       }
 \def\LSAP#1#2{\Bigl({2\over 3}\phip\ddpp{}#1\pp
 -{4\over 3}#1\pp +#2\pp \Bigr)}
 \def\II#1#2{\intl_0^1\Bigl\{G_x\ppp\LSAP{#1}{#2}\Bigr\}d\phip}
 \def\hk{{\hat k}}
 \def\hm{{\hat m}}
 \def\hX{{\hat X}}
 \def\of{\overline{f}}
 \def\og{\overline{g}}
\def\De{\delta\!e}
\def\DE{\delta\!E}
\def\DA{\delta\! A}
\def\DeS{\Delta \Sigma }
\def\DS{\delta\!\Sigma}
\def\DDS{\delta^2\!\Sigma}
\def\Ds{\delta\!S}
\def\DDs{\delta^2\!S}
\def\Dsth{\delta\!S_{\rm th}}
\def\Deps{\delta\!\eps}
\def\Dth{\delta\!\theta_{\rm th}}
 \def\hmass{m_{\scriptstyle {\rm H}}}
 \def\mui{\mu_{\scriptstyle {\rm I}}}
 \def\rhop{\rho^\prime}
 \def\vvec{{\bf v}}
  \def\vquot{{v_<^l\over v_>^{l+1}}}
 \def\3{\ss }
 \def\intl{\int\limits}
 \newcount\notenumber  \notenumber=0
 \def\note{\advance\notenumber by1 \footnote{$^{\the\notenumber}$}}
 \def\value#1#2#3#4{$#1\!=\!#2 \cdot\!10^{#3}{\rm #4}$}
 \def\val#1#2#3{$#1\!=\!#2{\rm #3}$}
 \def\ravmin{\overline{r_{\rm min}}}
 \def\urg{u_{\rm relg}}
 \def\urs{u_{\rm rel*}}
 \def\scoll{\sigma_{\rm coll}}
 \def\msol{{\rm M}_{\odot}}
 \def\pc{{\rm pc}}
 \def\yr{{\rm yr}}
 \def\mom#1{<\!#1\!>}
 \def\erf{{\rm erf}}
 \def\eps{\varepsilon}
 \def\Dvp{<\Delta {\dot v}_{\parallel}>}
 \def\Dvr{<\Delta {\dot v}_r>}
 \def\Dvrs{<\Delta {\dot v}^2_r>}
 \def\Dvps{<\Delta {\dot v}^2_{\parallel}>}
 \def\Euv{\langle E_{uv} \rangle }
 \def\vper{v_{\perp}}
 \def\vpar{v_{\parallel}}
 \def\L#1{\leftline{#1}}
 \def\phip{\phi^{\prime}}
 \def\parder#1#2{{\partial #1\over\partial #2}}
 \def\div#1{{1\over r^2}\parder{}{r}\left(r^2 #1 \right)}
 \def\ddt#1{{\partial #1\over\partial t}}
 \def\dedet#1#2{\left(\delta #1\over \delta t\right)_{\rm #2}}
 \def\rhos{\rho_*}
 \def\us{u_*}
 \def\rhog{\rho_g}
 \def\ug{u_g}
 \def\sigr{\sigma_r}
 \def\sigt{\sigma_t}
 \def\sigT{\sigma_T}
 \def\sig{\sigma}
 \def\centerline#1{\line{\hss#1\hss}}
 \def\h#1{{\hat #1}}
 \def\t#1{{\tilde #1}}
 \def\d#1{{\dot #1}}
 \def\oE{\overline{E}}
 \def\tf{{\tilde f}}
 \def\tg{{\tilde g}}
 \def\tB{{\tilde B}}
 \def\hB{{\hat B}}
 \def\sr{\sigma_r}
 \def\st{\sigma_{\theta}}
 \def\sp{\sigma_{\phi}}
 \def\vr{v_r}
 \def\vt{v_{\theta}}
 \def\vp{v_{\phi}}
 \def\Eq#1{(\the\ueberzaehla .#1)}
 \def\Fig#1{\the\ueberzaehla .#1}
 \def\s{\sigma}
 \def\o#1{\overline{#1}}
 \def\ui{u_{\infty}}
 \def\Ci{C_{\infty}}
 \def\ithrinf{\intl_{-\infty}^{\infty}\intl_{-\infty}^{\infty}
             \intl_{-\infty}^{\infty}}
 \def\cN{{\cal N}}
 \def\simgr{\,\raise 0.3ex\hbox{$>$}\kern -0.8em\lower 0.7ex
                 \hbox{$\sim $}\,}
 \def\simlt{\,\raise 0.3ex\hbox{$<$}\kern -0.8em\lower 0.7ex
                 \hbox{$\sim $}\,}
   \def\ApJ{Astrophys. J.}
   \def\ApJS{Astrophys. J. Suppl.}
   \def\AA{Astron. Astrophys.}
   \def\AAS{Astron. Astrophys. Suppl.}
   \def\MN{Monthly Notices Roy. Astron. Soc.}
   \def\AJ{Astron. J.}
   \def\Nat{Nature}
   \def\ASS{Astrophys. Space Sci.}
   \def\SSR{Space Sci. Rev.}
   \def\AL{Astr. Letters}
   \def\SA{Soviet Astron.}
   \def\ARAA{Ann. Rev. Astron. Astrophys.}
   \def\Mitt{Mitt. d. Astron. Ges.}
   \def\ESOM{ESO Messenger}
   \def\SAL{Sov. Astr. Lett.}
   \def\PhD{Phys. Rev. D}
   \def\Prog{Prog. theor. Phys.}
   \def\PASJ{Publ. astr. Soc. Jap.}
   \def\Acta{Acta astr.}
   \def\CPC{Comp. Phys. Comm.}
   \def\Bull{Bull. Am. Astr. Soc.}
   \def\PASP{Publ. astr. Soc. Pac.}
   \def\ZA{Zeits. f. Astroph.}
   \def\AGAS{Astron. Ges. Abstr. Ser.}
   \def\JRAS{J. R. astr. Soc. Can.}
 \def\usrref#1#2#3#4#5{\ref #2: #3, {\it #1}~{\bf #4}, #5}

 \def\QQ{\qquad}
 \newdimen\zlgfollower
 %
 \newcount\jourz
 \newcount\jourzmax \jourzmax=27
 \newtoks\journal
 \newbox\quadbox
 \def\ApJ{ApJ} \def\MN{MN} \def\ApJS{ApJS} \def\AA{AA} \def\AAS{AAS}
 \def\AAL{AAL} \def\AJ{AJ} \def\Nat{Nat} \def\ASS{ASS} \def\SSR{SSR}
 \def\AL{AL} \def\SA{SA} \def\ARAA{ARAA} \def\Mitt{Mitt} \def\ESOM{ESOM}
 \def\SAL{SAL} \def\PhRD{PhRD} \def\Prog{Prog} \def\PASJ{PASJ}
 \def\Acta{Acta} \def\CPC{CPC} \def\Bull{Bull} \def\PASP{PASP}
 \def\ApJL{ApJL} \def\Text{Text} \def\ZAph{ZAph} \def\AGA{AGA}
 \def\AP{AP}
 \def\REF#1#2#3#4#5#6{%
 \ifx#1\Text \jourz=0 \fi
 \ifx#1\ApJ \jourz=1 \fi\ifx#1\ApJL \jourz=2 \fi\ifx#1\ApJS \jourz=3 \fi
 \ifx#1\AA \jourz=4 \fi \ifx#1\AAL \jourz=5 \fi \ifx#1\AAS \jourz=6 \fi
 \ifx#1\MN \jourz=7 \fi \ifx#1\AJ \jourz=8 \fi \ifx#1\Nat \jourz=9 \fi
 \ifx#1\ASS \jourz=10 \fi\ifx#1\SSR \jourz=11 \fi\ifx#1\AL \jourz=12 \fi
 \ifx#1\SA \jourz=13 \fi \ifx#1\ARAA \jourz=14 \fi
 \ifx#1\Mitt \jourz=15 \fi \ifx#1\ESOM \jourz=16 \fi
 \ifx#1\SAL \jourz=17 \fi \ifx#1\PhRD \jourz=18 \fi
 \ifx#1\Prog \jourz=19 \fi \ifx#1\PASJ \jourz=20\fi
 \ifx#1\Acta \jourz=21 \fi \ifx#1\CPC \jourz=22 \fi
 \ifx#1\Bull \jourz=23 \fi \ifx#1\PASP \jourz=24\fi
 \ifx#1\ZAph \jourz=25 \fi \ifx#1\AGA \jourz=26\fi
 \ifx#1\AP  \jourz=27 \fi

  \ifnum\jourz<1
   \zlgfollower=\hsize
   \setbox\quadbox=\hbox{\QQ}
   \advance \zlgfollower by -\wd\quadbox
   \parindent=0.pt 
   \hyphenpenalty=5000 \parshape 2 0.pt
   \hsize \wd\quadbox \zlgfollower
   {{\rm #2} \par}
  \else
   \ifnum\jourz=1 \journal={ApJ}                    \fi
   \ifnum\jourz=2 \journal={ApJ}                    \fi
   \ifnum\jourz=3 \journal={ApJS}             \fi
   \ifnum\jourz=4 \journal={A\&A}               \fi
   \ifnum\jourz=5 \journal={A\&AL}     \fi
   \ifnum\jourz=6 \journal={A\&AS}        \fi
   \ifnum\jourz=7 \journal={MNRAS}\fi
   \ifnum\jourz=8 \journal={AJ}                       \fi
   \ifnum\jourz=9 \journal={Nat}                           \fi
   \ifnum\jourz=10 \journal={Ap\&SS}           \fi
   \ifnum\jourz=11 \journal={Space Sci. Rev.}                 \fi
   \ifnum\jourz=12 \journal={Astr. Letters}                   \fi
   \ifnum\jourz=13 \journal={SvA}                  \fi
   \ifnum\jourz=14 \journal={ARA\&A}    \fi
   \ifnum\jourz=15 \journal={Mitt. Astron. Ges.}           \fi
   \ifnum\jourz=16 \journal={ESO Messenger}                   \fi
   \ifnum\jourz=17 \journal={SvAL}                \fi
   \ifnum\jourz=18 \journal={Phys. Rev. D}                    \fi
   \ifnum\jourz=19 \journal={Prog. theor. Phys.}              \fi
   \ifnum\jourz=20 \journal={PASJ}           \fi
   \ifnum\jourz=21 \journal={Acta Astron.}                      \fi
   \ifnum\jourz=22 \journal={Comp. Phys. Comm.}               \fi
   \ifnum\jourz=23 \journal={Bull. Am. Astr. Soc.}            \fi
   \ifnum\jourz=24 \journal={PASP}           \fi
   \ifnum\jourz=25 \journal={Zeits. f. Astroph.}              \fi
   \ifnum\jourz=26 \journal={Astron. Ges. Abstr. Ser.}        \fi
   \ifnum\jourz=27 \journal={Afz }                    \fi
   \ifnum\jourz>\jourzmax
     \message{** no journal stored under this number, Ap.J. assumed **}
     \journal={Ap.J.}
   \fi
   \zlgfollower=\hsize
   \setbox\quadbox=\hbox{\QQ}
   \advance \zlgfollower by -\wd\quadbox \noindent
   \hyphenpenalty=5000 \parshape 2 0.pt
   \hsize \wd\quadbox \zlgfollower %
   \noindent
  {{#2, #3, \allowbreak { \the\journal},~{#4}, #5 #6} \par }
  \fi }

 \def\pn{\par\noindent}
 \def\sn{\smallskip\noindent}
 \def\mn{\medskip\noindent}
 \def\bn{\bigskip\noindent}
 \lineskip =2.0 \lineskip
 \centerline{\bf COMPARING DIRECT N-BODY INTEGRATION }\pn
 \centerline{\bf WITH ANISOTROPIC GASEOUS MODELS OF STAR CLUSTERS}
 \bigskip
 \bigskip
 \centerline{ Mirek Giersz
 \footnote{$\!{}^1$}{\it on leave from:
 Nicolaus Copernicus Astronomical Centre, Bartycka 12,
 OO-716 Warsaw, Poland},
 Rainer Spurzem
 \footnote{$\!{}^2$}{\it on leave from:
 Institut f"ur Theoretische Physik und Sternwarte, Olshausenstr. 40,
 D-W-2300 Kiel, Germany}}
 \bigskip
 \centerline{Department of Mathematics and Statistics,}
 \centerline{University of Edinburgh,}
 \centerline{King's Buildings,}
 \centerline{Edinburgh EH9 3JZ,}
 \centerline{U.K.}
 \bigskip
 \leftline{\bf ABSTRACT}
 \smallskip
 We compare the results for the dynamical evolution of star clusters
 derived from anisotropic gaseous models with the data from
 $N$-body simulations of
 isolated and one-component systems, each having
 modest number of stars. The statistical quality of $N$-body data was
 improved by averaging results from many $N$-body runs, each with the same
 initial parameters but with different sequences of random numbers
 used to initialize positions and velocities of the particles. We
 study the development of anisotropy, the spatial evolution and energy
 generation by three-body binaries and its $N$-dependence. We estimate the
 following free parameters of anisotropic gaseous models: the time scale
 for collisional anisotropy decay and the coefficient in the formulae for
 energy generation by three-body binaries. To achieve a fair agreement
 between N-body and gaseous models for the core in pre- as well as
 in post-collapse only the energy generation by binaries had to be varied
 by N.
 We find that anisotropy has
 considerable influence on the spatial structure of the cluster particularly
 for the intermediate and outer regions.

 \bn

 \vfill\eject

 \leftline{\bf 1~~INTRODUCTION}
 \vskip11truept
One of the grand challenges of theoretical astrophysics is to
understand the dynamics of globular star clusters.  Aside from the
intrinsic interest of understanding the behaviour of large $N$-body
systems, the importance of the problem stems from its relation with
current research into the stellar content of globular star clusters.
(See, for example, many of the papers in Janes 1991).
Unfortunately the direct simulation of such rich stellar systems with
$N$-body modelling is not yet feasible (Hut, Makino \& McMillan 1988).
The gap between
the largest
useful computer models ($N \le 10000$) and the median globular star
cluster ($N\sim 5\times10^5$) can only be bridged at present by use of
theory. There are two main classes of theory (cf. Saslaw 1985):
(i) Fokker-Planck
models, which are based on the Boltzmann equation of the kinetic
theory of gases, and (ii) gas models, which can be thought of as a set
of moment equations of the Fokker-Planck model.

These simplified models are the only detailed models which are {\sl
directly} applicable to large systems such as globular star clusters.
But their simplicity stems from many approximations and assumptions
which are required in their formulation, and it is not clear how well
these correspond to the real systems in nature. The usual assumption
of spherical symmetry contradicts the asymmetry of galactic tidal
fields. Almost all Fokker-Planck calculations assume isotropy of the
velocity distribution, which contradicts the evidence of proper
motions and of model-building. The treatment of stellar escape in
Fokker-Planck models has long been problematic, and it is never
considered in gas models. Three- and four-body encounters can only be
modeled within our limited knowledge of the relevant scattering
cross-sections.

In view of all these uncertainties, the reliance which can be placed
on these simplified models is doubtful. For this reason it is of
importance to test the predictions of these models against results
which are not subject to this wide class of simplifying assumptions
and approximations, i.e. $N$-body models. Despite the central role of
the simplified models in much recent research on cluster dynamics,
little has been done to test their validity. There is the already classical
comparison between the fluid-dynamical model of Larson (1970),
Monte-Carlo simulations, and direct $N$-body
integrations of 100 and 250 particles (Aarseth, H\'enon \& Wielen 1974,
see also Aarseth \& Lecar 1975), which were at that time maximal
particle numbers from the viewpoint of computational resources.

In the following years there has been much further improvement
of the theoretical
models as well as of computational resources. Direct numerical solutions
of the Fokker-Planck equation substituted the Monte-Carlo models,
and gaseous models with heat flux closure improved Larson's original
fluid-dynamical approach (Lynden-Bell \& Eggleton 1980, Heggie 1984).
Bettwieser \& Sugimoto (1985) tried to check the main assumption
of the gaseous model, the heat conductivity, by using a direct $N=1000$
model. Their $N$-body results, although in fair agreement with expectations
in pre-collapse, suffered from large statistical fluctuations especially
in post-collapse due to the still too small particle number.

With the advent of several large parallel computing facilities
as well as fast vector computers in general it is now
possible to extend direct $N$-body calculations in two respects: first
to improve the statistics for rather low $N$ (up to a few thousand)
by computing many independent models simultaneously, one on each
processor, improving the statistics then by averaging. On the other hand
the use of one of the fastest available vector supercomputers (CRAY YMP)
yielded for the first time high accuracy
$N$-body models for as much as 10000 particles throughout most of the
core collapse phase
(Spurzem \& Aarseth 1993).

This paper is designed to complement a larger survey of star cluster
evolution (Giersz \& Heggie 1993ab), which compares $N$-body models with
isotropic gaseous and Fokker-Planck results, by one particular aspect;
this is to check the reliability of a generalized gaseous model which
includes the possible generation of anisotropy in the radial and tangential
dispersions of the star cluster. On the other side there is a feedback
{}from the $N$-body models back to some adjustments in the gaseous model
code in order to reach optimal agreement of both models.

In the following section theory and numerical solutions of
anisotropic gaseous models are elaborated. Sect. 3 presents some
information on the $N$-body models used, to an extent that is necessary
here for understanding this paper; more details will be given elsewhere
(Giersz \& Heggie 1993a). Sect. 4 gives an account of the results,
which are finally discussed and complemented by concluding remarks
in the final section.
\vskip22truept
 \leftline{\bf 2~~ANISOTROPIC GASEOUS MODELS OF STAR CLUSTERS}
 \vskip11truept
 \leftline{\bf 2.1~~The model}
 \vskip11truept
 Observational fits of globular clusters (cf. e.g. Lupton \& Gunn 1987,
 Lupton, Gunn \& Griffin 1987)
 and direct $N$-body calculations show that there is a considerable
 amount of anisotropy ($\sr^2 > \s_t^2 $) in their halo. Including
 the effects of anisotropy into models of the dynamical cluster evolution
 has posed some difficulties in the past. The only 2-D orbit average
 Fokker-Planck models
 for anisotropic single mass clusters with and without a massive central
 object
 (Cohn \& Kulsrud 1978, Cohn 1979, Cohn 1985) have not been used further and
 there were a variety of anisotropic gaseous models with different
 closures (Bettwieser 1983, Bettwieser \& Spurzem 1986);
 however their relation to real many-body systems and the
 quality of the numerical solutions remained unclear. Recently, however,
 self-similar anisotropic models for regular pre-collapse as
 well as for singular post-collapse clusters were obtained
 (Louis \& Spurzem 1991, henceforth LS). This is an occasion to revisit
 the quality of the numerical solutions of the anisotropic gaseous model
 and to utilize such a model further to compare its predictions
 with the results of direct $N$-body calculations, with emphasis on
 the results concerning the anisotropy.
 \pn
 For the sake of completeness and to clarify small differences in
 the model and notation compared with LS
 we will present here the full set of equations
 used; dependent variables are the mass $M_r$ contained in a
 sphere of radius $r$, the local mass density $\rho $, radial
 and tangential pressure $p_r$, $p_t$, bulk mass transport velocity
 $u$, and transport velocities $v_r$, $v_t$ of the radial and tangential
 energy, respectively. As auxiliary quantities we use the radial
 and tangential 1-D velocity dispersions $\sr^2 = p_r/\rho $,
 $\s_t^2 = p_t/\rho $, the average velocity dispersion
 $\sigma^2 = (\sr^2 + 2\s_t^2)/3 $,
 the anisotropy $A=2-2\s_t^2/\sr^2 $ (note that $A=6a/(1+2a)$,
 where $a$ is the anisotropy measure used in LS),
 and the relaxation time
 $$T = {9\over 16 \sqrt{\pi}} {\sigma^3 \over G^2 m \rho \log(\gamma N)}
    \eqno(1) $$
 in the definition of Larson (1970),
 where $N$ is the total particle number of the star cluster,
 $m$ the individual stellar mass and
 $\gamma $ a numerical constant whose value will be discussed below.
 The equations are
 $$ \parder{M_r}{r} = 4 \pi r^2 \rho   \eqno(2) $$
 $$ \parder{\rho}{t} + {1\over r^2}\parder{}{r}(\rho u r^2 ) = 0
                                       \eqno(3) $$
 $$ \parder{u}{t}+u\parder{u}{r} + {GM_r\over r^2} +
     {1\over\rho}\parder{p_r}{r} + 2{p_r - p_t\over\rho r} = 0 \eqno(4) $$
 $$ \parder{p_r}{t} + {1\over r^2}\parder{}{r}(p_r u r^2 ) +
     2 p_r \parder{u}{r} + {3\over r^2}\parder{}{r}(p_r (v_r-u) r^2 ) -
      4 {p_t (v_t-u)\over r} =
     -{2\over 3} {p_r - p_t \over \lambda_{\rm A} T_{\rm A} } +
      \dedet{p_r}{\rm bin3}   \eqno(5) $$
 $$ \parder{p_t}{t} + {1\over r^2}\parder{}{r}(p_t u r^2 ) +
     2 {p_t u\over r} +{1\over r^2}\parder{}{r}(p_t (v_t-u) r^2 ) +
     2 {p_t (v_t-u)\over r} =
      {1\over 3} {p_r - p_t \over \lambda_{\rm A} T_{\rm A} } +
      \dedet{p_t}{\rm bin3}   \eqno(6) $$
 $$ v_r - u + {\lambda\over 4\pi G \rho T} \parder{\sigma^2}{r} = 0
                              \eqno(7) $$
 $$ v_r = v_t       \eqno(8) $$
 They are nearly equivalent to
 model A of LS (equal
 velocities for the transport of radial and tangential thermal
 energy). Note
 that model A (the so-called ``1 flux'' model) is very similar
 to the gaseous model of Bettwieser \& Spurzem
 (1986), whereas model B of LS
 (``2 flux'' model) is related to that of
 Bettwieser (1983). The net transport velocities for radial and
 tangential energy $(v_r-u)$ and $(v_t - u)$ can be derived from
 the energy fluxes $F_r$ and $F_t$ used as variables in these papers
 by dividing out
 a convenient multiple of the relevant pressure
 ($2p_t$ for $(v_t-u)$, $3p_r$ for $(v_r-u)\ $).
 The reader interested in more details about this and the
 connection of the variables to moments of the stellar velocity
 distribution is referred to LS.
 \pn
 Apart from the energy generation term due to hard binaries discussed
 below there is one more difference to LS in keeping the hydrodynamical
 terms in Eq. (4); for all applications comparing with $N$-body calculations
 this did not make any differences, however the stability of the code
 at very small timesteps and high central densities
 (near core bounce) is enhanced; if one follows
 core collapse over many orders of magnitude increase in central density
 these terms slightly affect the value of the anisotropy in comparison
 with the self-similar models of LS. Hence they were omitted only for test
 calculations to be compared quantitatively with the results of that
 paper.
 \par
 The numerical constants $\lambda_{\rm A} $ and $\lambda $ occurring
 in Eqs. (5) to (7) are related
 to the timescales of collisional anisotropy decay and heat transport,
 respectively. $\lambda $ is related to the standard
 $C$ constant in isotropic gaseous models (see e.g. Heggie \& Stephenson
 1988) by
 $$\lambda = {27\sqrt{\pi}\over 10} C   \eqno(8) $$
 $T_{\rm A}$ is the anisotropy decay timescale for an
 anisotropic local velocity distribution function; in
 the Appendix it is outlined how one gets the final result
 $T_{\rm A} = 10 \,T /9$, provided a particular velocity
 distribution function is assumed.
 Since the real distribution function is not completely known within
 the framework of a gaseous model
 a free numerical constant $\lambda_{\rm A} $ is
 introduced in Eqs. (5) and (6);
 the values chosen for $\lambda_{\rm A} $ and $\lambda $ will be discussed
 in comparison with the direct $N$-body calculations later. Note that
 in LS and all other previous papers
 the timescale $T_{\rm A} $ for an isotropic stellar background
 distribution as derived by Larson (1970) was used, which is
 $5\,T/6 $. Thus all models of LS correspond to a
 choice of $\lambda_{\rm A}=3/4$ here.
 \pn
 For the following comparison $N$-body models consisting of equal
 gravitating point masses are chosen, where any effects of finite
 size stars or stellar evolution are neglected. In that case the
 dominant energy source, which finally halts core collapse is
 the energy generated by formation and hardening of three-body
 binaries, which we take in accord with the standard ansatz
 (Goodman 1987) as
 \goodbreak\bigskip\goodbreak
 $$\dedet{p_r}{bin3} = {2\over 3} C_b {\rho^3\over m\sigma^2}
                   \Bigl({Gm\over\sigma}\Bigr)^5
              \cdot \theta(t-t_{b0}) \eqno(9) $$
 $$\dedet{p_t}{bin3} = \dedet{p_r}{bin3}\ .  \eqno(10) $$
 This is an isotropic energy input. A $\theta$-function has been attached
 here to illustrate the possibility to start the binary energy generation
 not from the very beginning but at a time $t_{b0}$, which is close
 but not identical to the time of core collapse and will be
 elaborated in Sect. 4.
 \pn
 Although the standard ansatz
 chooses $C_b = 90$, we keep this value here as a free parameter
 for reasons which will also be discussed in Sect. 4.
 Such a type of energy source to
 model the average energy generation by three-body binary generation
 and hardening was first used by Bettwieser \& Sugimoto (1984) and
 Heggie (1984) for their isotropic gaseous models.
 \pn
 Note that our anisotropic gaseous model equations
 are identical up to second order to moment equations of the
 Boltzmann equation with a Fokker-Planck collisional term. The
 closure equations Eqs. (7) and (8) are related to the analogous
 equation of heat transfer in a gas, except that the timescale occurring
 in the conductivity here is the appropriate stellar dynamical relaxation
 time (Lynden-Bell \& Eggleton 1980); thus such models are
 denoted as gaseous models to distinguish them from higher order
 fluid dynamical models (e.g. Louis 1990).
 \pn
 Altogether the anisotropic gaseous model itself contains three parameters,
 $\lambda $, $\gamma $, $\lambda_{\rm A}$; including the binary
 energy generation adds another two parameters, namely $C_b$ and
 $t_{b0}$.
 It is shown in a separate paper
 comparing isotropic gaseous models with direct $N$-body calculations
 (Giersz \& Heggie 1993a) that for all models (including also
 direct solutions of the Fokker-Planck equation) values close to
 $\gamma = 0.11$ and $C=0.104$ (i.e. $\lambda = 0.4977 $),
 give the best agreement; the isotropic and anisotropic gaseous
 models differ only marginally in the evolution of the inner mass shells
 during pre- and post-collapse evolution; this is shown in Fig. 1,
 where an isotropic model obtained with Heggie's code is
 compared with an anisotropic 1 flux model of Spurzem's code.
 Since the behaviour of the inner Lagrangian radii determines
 the best values for $\gamma $ and $C$ as discussed by Giersz \&
 Heggie (1993a) we use here just the same values.
 \mn
 It is the focus of this paper
 to show how
 a comparison with direct $N$-body calculations enables us to find
 unambiguously an optimal set of the remaining parameters
 $\lambda_{\rm A}$, $C_b$, and $t_{b0}$ in order to
 achieve fair agreement
 with the $N$-body results.
 \pn
 Without binary energy generation the quasi-static evolution of a
 system governed by Eqs. (2) to (8) depends only on the product
 $\lambda\cdot\lambda_{\rm A}$; different choices for the total
 particle number result merely in a rescaling of time. With
 binary energy generation, however, increasing
 $N$ means a smaller binary energy generation;
 thus the post-collapse evolution is qualitatively different
 for varying particle number, ranging from simply steady reexpansion
 for low $N$, through periodic oscillations at intermediate $N$ and
 probably chaotic large-amplitude gravothermal oscillations
 (originally detected by the isotropic gaseous model of Bettwieser
 \& Sugimoto, 1984) at
 very large $N$
 (Heggie \& Ramamani 1989, Cohn, Hut \& Wise 1989).
 \vskip22truept
 \leftline{\bf 2.2~~Numerical Solution of the Anisotropic Gaseous Model
 Equations }
 \vskip11truept
 For the numerical solution of the model equations and comparisons
 with direct $N$-body results standard $N$-body units were used, where
 $G=1$, the total mass of the system $M=1$, and the total energy
 of Plummer's model, which was used as initial model, is $E=-1/4 $;
 in these units the scaling radius of Plummer's model is $a= 3 \pi/16 $.
 The equations were discretized on an Eulerian mesh with 200
 logarithmically equidistant grid points
 (for some test runs 400); they were distributed between a minimal and
 maximal radius of $2.06 \cdot 10^{-6} $ and $144$ in the above
 units.
 \pn
 The equations used for discretization and numerical solution
 are equivalent but not identical to those of Eqs. (2) to (8). First
 instead of the physical quantities we used
 $\log M_r $, $\log\rho $, $\log p_r $ and $\log p_t $ as
 well as the {\sl net} energy transport velocities $v_r-u$ and $v_t-u$
 together with $u$ as dependent variables; in the Eulerian mesh scheme
 $r$ is the independent variable.
 In order to achieve the utmost accuracy
 of mass and energy conservation a partially implicit scheme was used
 where the equations were formulated in terms of $\zeta X +
 (1-\zeta) X^\star $, where X is the actual value and $X^\star $ the
 one of the previous timestep; such a scheme is numerically unstable
 with the optimal value of $\zeta = 0.5 $, thus $\zeta = 0.55 $ was used.
 The difference equations were discretized on a logarithmically equidistant
 mesh and solved iteratively with a Newton-Raphson-Henyey method.
 For convenient discretization in the logarithmic variables the divergence
 terms occurring in Eqs. (3), (5), and (6) were all split up.
 As a criterion for choosing the timestep we used that the
 positive logarithmic quantities should not change by more than $0.05$,
 complemented by the condition that the timestep should not exceed the
 central relaxation time, which ensures good accuracy at the turning
 points from pre- to post-collapse. As boundary conditions we imposed
 time-independent logarithmic gradients of density and pressures at
 the outer, linear variation of the velocities with
 respect to $r$ and of the mass with respect to $r^3$ at the inner boundary.
 \pn
 The conservation of energy and mass in all results
 presented in the following section, related to the total values
 of $0.25$ and $1$, respectively, was always smaller then or equal to
 $0.25 \% $;
 although the strict mass conservation of earlier codes was lost
 (due to splitting the divergence term in Eq. 3 and the different
 boundary condition) the mass error remained small as stated; therewith,
 however,
 a much better quality of energy conservation was achieved than in
 previously published anisotropic
 models of Bettwieser \& Spurzem (1986) and Bettwieser (1983).
 Although they discretized the divergence terms in Eqs. (5) and
 (6) in a strictly conservative manner this did not ensure that
 the total energy, consisting of thermal and gravitational energy (and bulk
 kinetic energy, which is here not important) is strictly conserved; the
 energy conservation becomes worse due to the non-conservative form
 of terms if one has separate energy equations for radial (Eq. 5) and
 tangential energy (Eq. 6). On the contrary the logarithmic variables
 improve energy conservation since they vary linearly whenever there
 is a power law, which turned out to be the more important effect.
 \pn
 In fact it
 occurred that the achieved high level of accuracy was necessary here,
 because for the previous models the errors were larger
 than the remaining deviations between gaseous model and direct $N$-body
 calculation.
 \pn
 The numerical method was tested by comparing with the known self-similar
 solutions of LS; for that purpose the spatial resolution was enhanced
 by choosing 400 grid points with an innermost radius of $1.11\cdot 10^{-8}$
 (to be comparable with LS the two first hydrodynamic terms in Eq. (4)
 had been omitted for these calculations only).
 A pure pre-collapse solution was followed (without binary
 energy generation)
 over an increase in central density of 13 orders of magnitude;
 the energy error here remained below $1\% $ until the density increased
 roughly 8 orders of magnitude - thereafter the resolution of the
 core became poorer and consequently the energy error increased further.
 It would be possible to increase the accuracy even after such a
 large growth of central density by either more meshpoints or
 moving them with e.g. the shrinking core radius. This
 is not useful for our purpose here, since we do not follow the
 system deep into core collapse.

\mn
 Figs. 2ab show the
 radial profiles of the logarithmic density power law index $\alpha $
 and of the anisotropy $A$ for a number of models with
 different evolutionary time; the onset of the self-similar
 collapse phase is clearly visible. In Figs. 3ab
 the nearly constant values of $\alpha $ and $A$
 after the self-similar phase has been entered are shown as a function of
 $\lambda_{\rm A}\lambda $ in
 comparison to their values in the self-similar anisotropic
 pre-collapse models of LS, to the isotropic gaseous models of
 Lynden-Bell \& Eggleton (1980), and to the direct solution of
 the isotropic Fokker-Planck equation (Cohn 1980).
 (Since LS only published two different $\lambda $
 parameters, additional results for different $\lambda_{\rm A}\lambda $
 have been obtained by P.D. Louis, priv. communication). Whereas in
 LS $\lambda $ was adjusted to match the
 asymptotic core collapse rate with the results of higher order
 fluid dynamical models ($\lambda = 0.186 $, Louis 1990), we chose here a
 value
 of $\lambda = 0.4977$ (equivalent to $C=0.104$), which yields fair
 agreement with Fokker-Planck results (Heggie \& Stephenson 1988) as
 well as with our $N$-body results in pre- and post-collapse. Throughout
 most of our calculations we chose $\lambda_{\rm A}=0.1$, for reasons
 which will become clear in Sect. 4;
 a few runs were performed to show the
 effect of choosing
 $\lambda_{\rm A}=1$.
 \mn
 For our comparisons presented here we use only rather low particle
 numbers ($N=250$, $500$, $1000$, $2000$); the post-collapse solution in such
 cases is a steadily reexpanding system with a regular
 core, related to the stable
 post-collapse isotropic self-similar models of Goodman (1987).
 A survey of larger particle numbers with the
 anisotropic gaseous model presented here where
 the post-collapse model is unstable to gravothermal oscillations is
 in progress (Spurzem \& Louis 1993). Occasionally we will refer
 to another comparison of this model with an $N=10000$ direct
 $N$-body calculation (Spurzem \& Aarseth 1993).
 \vskip22truept
\leftline{\bf 3~~$N$-BODY CALCULATIONS}
\vskip11truept
Over the last few years a new kind of computer
became available for scientific use,
consisting of a set of parallel transputers or vector processors.
Their use
opens a new chapter for stellar dynamical study. The power of these
computers can be exploited in two different ways. One direction is simply
to carry out calculations more quickly. This means to study larger systems.
A second way is to obtain results having better statistics.
It is well known that
errors of the positions and velocities of the stars grow exponentially on
a time scale shorter than the crossing time
(Miller 1964, Goodman, Heggie \& Hut 1993).
Nevertheless $N$-body modeling practitioners believe that
statistical results obtained from N-body calculations
are meaningful (cf. e.g. Aarseth \& Lecar 1975).
Therefore it is worth paying much more attention to improving the
statistical
quality of such models.
This aim was achieved by running the same model several times in parallel,
where the individual processes
differ only with respect to the randomly generated initial positions and
velocities of the stars. A full discussion of the method and the hardware
used in the simulations of evolution of small $N$-body systems
($N = 250$, 500,
1000, 2000) is described in great detail in the separate paper (Giersz
\& Heggie 1993a). Here we only outline them.
\pn
The computational work has been carried out on two different parallel
computers installed at the Edinburgh Parallel Computing Centre. One is the
Meiko transputer array consisting of 400 processors, the
other is the Meiko vector processor array containing 64 processors. Models
consisting of $N = 250$, 500 and 2000 particles were computed on the
transputer
array using 56, 56 and 16 processors, respectively. The model for $N= 1000$
bodies
was computed on the vector array using 40 processors, and on the DEC Alpha
superclaster machine (called ``fringe'') for 20 cases.
\mn
The first sets of models ($N = 250$, 500, 1000) were computed using a code
developed mainly by Heggie (Heggie 1973). The $N=2000$ model, not fully
completed yet, and part of the $N=1000$ model were
computed using a version of Aarseth's standard code NBODY5
(Aarseth 1985), as well as the $N=10000$ case, which will be presented
in more detail elsewhere (Spurzem \& Aarseth 1993).
The initial conditions for these models were drawn from Plummer's model
(isolated system), with equal masses and no primordial binaries.
In intervals of one $N$-body time unit there was output produced for each
of a parallel set of cases.
This output mainly consisted of the following information: 1) - Lagrangian
radii (1\%, 2\%, 5\%, 10\%, 20\%, 50\%, 75\%) with respect
to the centre of density,
2) -- the mean square radial and tangential velocities for each Lagrangian
shell, 3) -- global information: number of bound stars,
energy of the system, number
and energy of escapers, 4) -- information about binaries: radius, internal
energy, number and energy of escapers.
The output from all models (different processors) was collected in a
single file and later analyzed (averaged over all models)  to produce
statistics for all important quantities at each time.
\mn
The improvement in the statistical quality of the data achieved in our
simulations allows us to perform very detailed comparisons with other
methods, as in this case with the anisotropic gaseous model.

 \vskip22truept
 \leftline{\bf 4~~RESULTS}
 \vskip11truept

 Three different groups of complete
 $N$-body runs, consisting of $N=250$, $500$, and $1000$ particles,
 have been compared with the anisotropic gaseous models;
 two not fully completed $N$-body models will also be partly discussed
 ($N=2000$, $N=10000$). In what follows a phrase such as
 ``1000-body model'' refers to values averaged over many models,
 except for $N=10000$ where there is only one case yet available.
 \bn
 We compare Lagrangian radii containing certain fractions of the
 total mass as a function of time and the anisotropy of the
 velocity distributions averaged within these Lagrangian mass
 shells as a function of time for both the gaseous and $N$-body models.
 All $N$-body data had to be adjusted in their initial Lagrangian
 radii since the construction of an initial $N$-body model, its scaling
 to the required value of the initial potential energy, and the use
 of the density centre all result in the fact that the initial model is
 slightly biased with respect to Plummer's model;
 the gaseous model was initially much closer to it.
 Fig. 4 demonstrates the amount of shift necessary for the $N=1000$
 particle number as an example.
 \mn
 Fig. 5 depicts the evolution of the 1\% Lagrangian radius in the
 $N=1000$ model in comparison to gaseous model results with a
 varying initial time $t_{b0}$ for the binary energy generation.
 The value of $t_{b0}$ controls the time at which the gaseous model
 curve deviates from a pure core collapse case without any energy
 generation. It does not alter very much the post-collapse behaviour,
 but we selected for the three particle numbers $250$, $500$, and
 $1000$ values of $t_{b0}=50$, $130$, and $230$, respectively, which
 appeared as best fits of the gaseous model curves to the
 $N$-body case. Note that the best fit is usually meant with respect to
 the $5\%$, $2\%$ and $1\%$ Lagrangian mass shells. We have
 chosen those radii because they are only slightly smaller than the core
 radius, within which most of the binary activity occurs. The
 evolution of the outer shells does not depend strongly on
 $t_{b0}$.
 \mn
 Physically $t_{b0}$ should
 be related to a time at which binary activity begins to play
 a role in the $N$-body system; the ratios of the total binary energy
 released in the N-body system
 until the time $t_{b0}$ to the initial total energy of the system, however,
 were not the same for all particle numbers (the fractions were
 $2.48\%$, $2.05\%$, $0.13\%$ for $N=250$, $500$, $1000$, respectively,
 see discussion of $x$-values below and in Sect. 5).
 \mn
 Another parameter is the strength of the binary energy generation $C_b$;
 Fig. 6a shows that the minimum of the curve for the innermost
 Lagrangian radius (1\%) is clearly of a different shape for varying
 $C_b$-values; for smaller $C_b$ the minimum is less shallow, because
 core collapse can proceed further to higher densities, until the
 energy generation suffices to halt and reverse it. There is also
 a variation in the total energy input for different $C_b$ which can
 be recognized by the different post-collapse curves. Again for each
 particle number we could select those values of $C_b=55$, $70$, and
 $90$, for $N=250$, $500$, $1000$, respectively, which appeared to fit
 best the shape of
 the minimum of the $N$-body result as well as its post-collapse slope.
 Since there are always some
 fluctuations in the $N$-body system we varied $C_b$ only in larger
 steps ($45$, $55$, $70$, $90$, $135$) without attempting to determine
 the value more exactly. Moreover, one should note that due to a loss
 of cases in the parallel $N$-body calculations (for some individual
 models the total energy was not preserved due to very strong
 interactions between binaries and field stars and between binaries
 themselves (Giersz \& Heggie 1993a)) the statistics of the
 corresponding averaged model worsen towards the end of the presented
 post-collapse evolution. For $N=250$, $500$, $1000$ at the time $t=600$
 up to half of the initial number of cases were lost.
 \mn
 According to Goodman (1987) (see his Eq. II.14)
 $C_b$ measures the product of the quantities
 $x=3\phi_c / v^2_c $, where $\phi_c$ and $v_c$ are the central
 potential and 3-D velocity dispersion, and
 $f_3$, which describes the amount of energy supplied to the core
 by an individual binary.
 \mn
 Fig. 6b depicts the evolution of $x$ for different $N$-body models.
 The time for all models $N\ne 1000$ was
 rescaled such that the distant two-body encounter relaxation
 timescale has the same value as for $N=1000$;
 (this leads as one should
 expect to very good agreement of all pre-collapse
 curves). The
 curves were obtained by smoothing the original data using the standard
 procedure SMOOFT from Numerical Recipes (Press et al. 1986).
 \mn
 Note that we find for $N=1000$ a value
 of $C_b=90$, which is expected if the average total energy available
 {}from a binary resident in the core is supplied to the core ($f_3\approx 1$).
 For lower particle numbers the changes of $x$ with $N$ during the
 post-collapse phase cannot be only accountable for the estimated values
 of $C_b$. This already
 suggests that for these models only a fraction, $f_3$, of
 the energy released by 3-body binaries is deposited into the core,
 $f_3\approx 0.9$ for $N=500$ and $f_3\approx 0.8$ for $N=250$,
 although we will give some more evidence on this point later.
 \mn
 The best values for $t_{b0} $ and $C_b$ were used
 to produce the data presented in Fig. 7 for the innermost Lagrangian
 radius ($N=250$: $C_b=55$, $t_{b0}=50$;
 $N=500$: $C_b=70$, $t_{b0}=130$; $N=1000$: $C_b=90$, $t_{b0}=230$),
 and in Figs. 8ab to 10ab for all three
 particle numbers showing the evolution of seven Lagrangian radii
 ranging from 1\% to 75\%. Looking at the innermost Lagrangian mass
 shells containing 1\%, 2\%, and 5\% of the total mass we conclude
 that there is an excellent agreement between both models. Such a
 fitting procedure with the anisotropic gaseous model
 as described above would not be
 possible if the $N$-body results were not improved in their statistics
 by computing multiple cases in parallel, which suppresses the intrinsic
 noise of the individual $N$-body system.
 \mn
 Note, however, that for $N=250$ the best fit to the
 post-collapse expansion made by $C_b=55$
 for the inner Lagrangian mass shells did not automatically
 produce the correct form of the minimum at core bounce
 in contrast to the other particle numbers $N=500$ and $N=1000$;
 for $N=250$ the minimum of the $N$-body model curves is always
 less shallow than for the gaseous model.
 \mn
 For intermediate
 Lagrangian radii containing 10\%,
 20\%, and 50\% of the total mass at $N=1000$ there is
 a tendency for the gaseous model to expand faster in post-collapse,
 at $N=500$ there is fair agreement between both models here, whereas
 for $N=250$ on the contrary the $N$-body models expand faster.
 The 75\% radius is in all $N$-body models further outside than
 in the gaseous model.
 The reasons for these remaining differences between our two
 models will be discussed also in Sect. 5.
 The larger expansion rate of the
 N-body system in the outer halo (75 \%) as compared with the gaseous model
 is connected both with substantial amounts of
 escapers (Goodman 1984) and with degradation of the statistics due to
 loss of cases. There is
 a different treatment of
 the outer boundary in both models;
 the $N$-body model
 had initially all particles within a sphere
 of radius 10 $r_h$ ($r_h$: half-mass radius) i.e. any formed in the initial
 Plummer model at larger radii were removed,
 in the course of the evolution, however,
 a significant fraction of very loosely bound halo particles
 and escapers move
 outwards and populate areas within a sphere of a radius
 much larger than the initial radius;
 this process starts from the very beginning with
 particles just moving outwards on radial orbits.
 The phase space is not completely covered in that region,
 e.g. the angular momentum of the particles there
 (which are approximately 5\% of all particles for
 $N=10000$ in the late collapse phase)
 is always much less than the maximum angular momentum
 of a circular orbit. On the contrary the gaseous model extended
 much further outwards already in the initial model in order to
 avoid dynamical
 perturbations originating from the boundary, but the system was
 assumed to be isotropic there initially, i.e. the accessible phase space
 was fully covered.
 Any expansion of the gaseous model's 75\% radius is a dynamical
 expansion driven by pressure against the outer shells, which is
 physically different to what happens in the $N$-body system,
 where some fraction of
 stars is just moving outwards filling empty space.
 Therefore the
 75 \% Lagrangian radius is always further outside in the $N$-body models.
 This effect is even more pronounced for the $90\%$ radius of the
 $N=2000$ body calculation as an example (Fig. 11), where one can recognize
 that the $N$-body model started with a smaller radius than the
 gaseous model, but expands from the beginning
 faster, since particles can move freely outwards in contrast to the
 gaseous model. The 90 \% radius has therefore not been shifted,
 since its difference with the gaseous model was much larger than usual.
 \mn
 We now start to discuss the evolution of the anisotropy
 $A= 2 - 2\s_t^2 /\sr^2 $; it is presented as a mass-weighted
 average taken over the spherical shells whose inner and outer radii
 are given by the Lagrangian radii of the previous figures.
 It is now the only remaining free parameter $\lambda_{\rm A}$
 which limits the anisotropy by collisional isotropization.
 Let us denote with $A_5$, $A_6$, $A_7$ the anisotropy
 between Lagrangian radii of $10 \% $ to $20 \% $, $20\% $ and
 $50 \% $, and $50\% $ to $100 \% $, respectively.
 Their time evolution is shown in comparison between $N$-body results
 and gaseous models for $N=250$, $500$, $1000$, $2000$, and $10000$ in
 Figs. (12) to (16), respectively. Most gaseous
 models were run with $\lambda_{\rm A}=0.1$
 (solid lines in Figs.), whereas in Figs. (12)-(14)
 alternative results for $\lambda_{\rm A}=1 $
 (dashed lines) are also presented.
 The anisotropy generation for the $\lambda_{\rm A}=1 $ case is clearly much
 too high. $\lambda_{\rm A}=0.1$, however,
 yields excellent agreement of
 $A_5$, $A_6$ and $A_7$ for $N=10000$ (Fig. 16), as well as for all other
 particle numbers until a certain time $t_{a0}$,
 which turns out to be approximately equal to $t_{b0}$ for
 $N=1000$, and a little smaller than $t_{b0}$ for $N=500$ and $N=250$.
 Since for the lower particle numbers $N<10000$
 we know that there has been
 already more binary activity at $t_{b0}$ it is consistent to
 postulate a connection between the binary activity in the core
 and the larger anisotropy (more radial energy) outside.
 \mn
 There is further evidence that the anisotropy in the outer shells
 is determined (at least partially) by binary activity. Firstly, there is
 a strong increase in the anisotropy after $t_{b0}$, when the binary
 energy generation starts, in all $\lambda_{\rm A}=1$ gaseous models
 (Figs. 12-14). For the $\lambda_{\rm A}=0.1$ case it is suppressed by stronger
 collisional isotropization. Secondly, the total energy and number of the
 bound hard binaries level off during the post-collapse evolution nearly
 at the same time as anisotropies do (Giersz \& Heggie 1993b).
 \mn
 To further clarify
 the point we performed an artificial experiment by substituting
 the density exponent of 3 in Eq. (9) with $3-\beta $ for
 the tangential and $3+\beta$ for the radial energy generation.
 We compared
 runs for $\beta = 1$ and $\beta = 0$, the latter being
 the standard case. The total energy generation
 was normalized such that it does not differ in both
 cases and such that it is still isotropic at the centre.
 For positive $\beta $ there is more radial binary energy
 generation than tangential; the physical picture is
 that encounters with binaries very likely produce a particle
 leaving the core with low angular momentum i.e. higher radial
 than tangential energy. It can be seen in Figs. (17) and
 (18) that this indeed is a good model for what happens with
 the anisotropy: the stronger radial energy generation does not
 alter $A_5$ and $A_6$ much, but there is a clear trend towards the
 $N$-body results for $A_7$.
 \mn
 As a last check we suppressed close encounter and binary activity
 in an $N$-body simulation using S. Aarseth's NBODY1 with a smoothing
 parameter
 $\varepsilon = p_0 $ and $\varepsilon = 2p_0$, where $p_0$ is the
 impact parameter for which the deflection in the relative orbit of
 interacting particles is equal to $\pi/2$. In $N$-body units
 $p_0=2/(NV^2)$, where for $V^2$ we use a value obtained from the virial
 theorem. Assuming that the maximum impact parameter is roughly equal
 to $r_h$ we can estimate for $N=250$ that for $\varepsilon = 0.016$
 and $\varepsilon = 0.032$ about $9\%$ and $21\%$ of all encounters
 is suppressed, respectively.
 \mn
 The striking
 result shown in Fig. (19)
 is the much better agreement between $N$-body and gaseous models
 for $A_7$ in the case $\varepsilon = 0.016 $; the trend is a little
 too strong for $\varepsilon = 0.032$. Taking this last result into
 account we conclude that $\lambda_{\rm A}=0.1$ seems to be the
 physically realistic value for all cases, provided close
 encounters and binary activity play little role; this is
 naturally the case in
 large $N$ systems (for which the gaseous models are qualified).
 It is consistent
 that in all results for $\lambda_{\rm A}=0.1$ and $t<t_{b0}$ the
 agreement is excellent; any difference starts for low $N$ and at
 times after which binary activity has started.
 \vskip22truept
 \leftline{\bf 5~~CONCLUSIONS AND DISCUSSION}
 \vskip11truept
 We have performed a set of direct $N$-body models
 of idealized single mass star clusters with improved statistics obtained
 by calculating independent models in parallel on a parallel
 computer for particle numbers of $N=250$, 500, 1000, and 2000. One
 model of $N=10000$ (Spurzem \& Aarseth 1993) complements some of our
 data. The results of the parallel $N$-body project are published
 in more detail
 in Giersz \& Heggie (1993ab); here
 we related them to time dependent evolutionary
 calculations with an anisotropic gaseous model, which is closely related
 to the models of Bettwieser \& Spurzem (1986) and
 Louis \& Spurzem (1991, LS),
 but complemented by an energy source appropriate to describe a
 (usually isotropic)
 energy input by formation and hardening of three-body binaries analogous
 to that used in the isotropic gaseous models of Bettwieser and Sugimoto
 (1984) and Heggie and Ramamani (1989).
 \pn
 Excellent agreement of the evolution of the innermost Lagrangian
 mass shells (less than or equal to the core radius) in both pre- and
 post-collapse could be found by adjusting the strength and start
 time of the binary activity in the gaseous model. Remarkably the
 numerical value of the binary strength $C_b=90$ found in that
 manner agrees very well for $N\ge 1000$ with theoretical expectations
 for systems whose evolution is dominated by small angle gravitational
 encounters (Goodman 1987).
 For $N=500$ and $N=250$ such agreement could be achieved
 only by choosing somewhat smaller values of $C_b = 70$ and $C_b = 55$,
 respectively.
 \mn
 The generation of anisotropy during pre-collapse and its levelling off
 in post-collapse within the half-mass radius
 is in fair agreement between $N$-body and gaseous models, provided
 one takes as timescale for collisional decay of anisotropy
 $\lambda_{\rm A}T_{\rm A}$, where $\lambda_{\rm A}=0.1$, and
 $T_{\rm A} $ is the local anisotropy decay time for an anisotropic
 Larson-type distribution function (see Eq. (5), (6),
 and (A15), and Larson 1970). For the outer portions of the system
 the agreement is satisfactory only if $N\ge 2000$ or time $t\le t_{b0}$, the
 latter being the time at which binary activity becomes important.
 \mn
 It is important to note that in pre-collapse and without binary
 activity we have scaled all $N$-body and gaseous models such that
 their collapse phases match exactly;
 the scaling factor was computed theoretically and the
 result is that all models for all particle numbers agreed excellently
 for a unique value of the remaining parameters ($\lambda = 0.4977$
 for heat conductivity, $\lambda_{\rm A} = 0.1$ for anisotropy
 decay, and $\gamma =0.11$ in the Coulomb logarithm, cf. Eqs. 2-8).
 The agreement shows up for the anisotropy everywhere in the system
 and for the evolution of the Lagrangian mass shells up to the
 core radius. It is a puzzle, that between the core and
 half-mass radius in pre- and post-collapse
 the evolution of the Lagrangian mass shells in the gaseous model
 does not agree very well with the
 $N$-body models.
 There is a tendency for the
 gaseous model to expand faster (this is also the case for the isotropic
 Fokker-Planck model, Giersz \& Heggie 1993b) at these radii as
 can be seen for the $N=1000$ case in Fig. (8ab); however the same
 result occurs in the runs available for the largest particle numbers
 $N=2000$ and $10000$ (no figures given in this paper).
Especially for $N=10000$ distant two-body
 encounters are the main force which drive the evolution. So the gaseous
 model based on the thermal conductivity approximation should in principle
 describe the evolution well. However, this is not the case, particularly
 for the $50\%$ Lagrangian radius. An explanation can be connected
 with the fact that the gaseous model cannot cope with non-local scattering.
 To mimic the energy transport in the radial and tangential directions
 we have to take into account all interactions along a particle orbit.
 This could be done for example by an orbit-averaged 2-D Fokker-Planck
 equation
 or by Monte-Carlo realization of this equation. We would like to stress that
 the best agreement with $N$-body data for the half mass radius is achieved by
 Stod\'o\l kiewicz's Monte-Carlo code (1982). Unfortunately any quantitative
 proof of this statement is not possible because the only obtainable data
 are diagrammatic.
 \mn
 For smaller systems ($N=250$ and $500$) that effect is present as well
 but we cannot see it because it is gradually overcome by
 the effect of reaction products of close two- and
 three-body interactions, which deposit their energy by large angle
 encounters in the outer regions of the system.
 \mn
 One may argue that a variation of $\gamma$, the
 factor in the Coulomb logarithm, or $\lambda $, the heat conductivity
 coefficient, should be checked to see whether it is sufficient
 for full agreement of the evolution of all Lagrangian
 mass shells.
 There is some theoretical evidence about a possible dependence of $\gamma$ on
 position and the total number of stars (see Spitzer 1987 for detailed
 discussion). Spitzer (1987) suggested that $\gamma$ could be equal to
 $2N_c/N$. From Heggie \& Stephenson (1988) one may conclude that
 $\lambda $ is slightly different between pre- and post-collapse.
 However, our results indicate that our adopted constants
 give very good agreement during the collapse phase. So the local
 effects of distant two-body encounters are correctly modeled in this phase.
 Discrepancies start to build up when the number of stars in the core
 is very small and binary activity starts to play a role. This can
 favour Spitzer's definition of $\gamma$, but on the other hand non-local
 effects connected with the small number of stars start to be increasingly
 important. For example
 very close two- and three-body scatterings can produce
 high velocity stars which would disturb the velocity
 distribution
 such that our second order moment equations are not sufficient to
 model the system. Our anisotropic gaseous model improves previous
 gaseous models with respect to the inclusion of (second order)
 anisotropy, but it may be too simplified in the presence of high-energy
 scattering events, which happen in the core predominantly for
 small particle numbers $N\le 1000$ in post-collapse. It would
 be interesting to check how an anisotropic gaseous model such as that
 of Louis (1990) or numerical solutions of the orbit-averaged
 2-D Fokker-Planck equation could cope
 with this problem. So we suggest that post-collapse evolution
 in principle can be understood
 without introducing new values of $\gamma$ and $\lambda$,
 although we cannot totally rule out that possibility (particularly for
 high $N$ systems) on the basis of our results.
 \mn
 There is more evidence that binary and close encounter activity
 causes differences between gaseous and $N$-body models for low $N$;
 this is the much too strong anisotropy generation outside of the
 half-mass radius in these cases (Figs. 12-14); first the problems
 occur as the binary activity and the anisotropy levels off in the $N$-body
 systems at the same time as the binary activity begins (Giersz \& Heggie
 1993b). Second, a test calculation with a local,
 but anisotropic (stronger radial) energy generation due to binaries
 exhibits a trend towards the $N$-body results. Third, we may interpret
 our empirically found strengths of binary energy generation
 in terms of the parameter $f_3$ of Goodman (1987) describing
 the fraction of binary energy liberated in the core. Our data
 indicate that $f_3 < 1$ for $N\le 500$. Since all binaries are
 located in the core this means a rapid non-local energy transfer
 to regions outside of the core radius, carried by particles moving
 outwards on radial orbits and creating the anisotropy in the halo.
 Fourth, a test N-body calculation with suppressed close encounters
 by choosing non zero smoothing parameter shows much better agreement
 with the gaseous model results.
 \mn
 Our findings are consistent with the results of earlier work of
 Spitzer \& Mathieu (1980) and Goodman (1984) stating that the
 reaction products of superelastic binary-single star scatterings will
 be transported outwards on elongated radial orbits, sometimes even
 escape. Angular momentum scattering along these orbits will distribute
 the energy of the reaction product in the corresponding radial zones.
 This is the mechanism by which the above mentioned non-local energy
 transport is provided and we identify the reaction products
 as carriers of the observed high anisotropy in
 the $N$-body models. It is not surprising that for systems with
 larger $N$ and higher $x$ (deeper central potential, see Fig. 6b)
 this energy distribution is confined more to the homogeneous core itself and
 thus more consistent with the gaseous model.
 \mn
 The clue to an understanding of the remaining further differences between
 the two models (sharp minimum of the Lagrangian radii,
 compare Figs. 7 and 10a, and higher
 binary energy generation for small $N$)
 lies in the nature of the binary and close encounter
 activity, whose character is very different for the low $N$ systems
 and for high $N$ systems.
 For the former there is a higher probability, with respect to the probability
 of small angle encounters, that core particles are
 subject to a close two-body encounter or a three-body binary will
 form. Therefore for $N=250$ binaries are created earlier and they can
 affect the core collapse and eventually stop it earlier
 (in terms of $x=3\phi_0/v_0^2$, see Fig. 6b)
 than in higher $N$ systems. Because of smaller $x$
 the rate of collapse
 is higher (Cohn 1980), and binaries have to
 generate more energy to influence
 core collapse and finally reverse it.
 Thereby we finally have found a natural explanation why the
 energy generated by binaries before the time $t_{b0}$ is higher. However,
 due to smaller $x$ (less deep central potential) a
 bigger fraction of the energy can be carried by particles non-locally
 outwards and it is not felt as early by the core as for higher
 particle numbers. This explains also why the intermediate shells for
 $N=250$ expand faster than in the gaseous model, because that is where
 much of the energy and mass is non-locally transported to.
 \pn
 A supplementary explanation for the sharper minimum in the time evolution of
 the inner Lagrangian radii in the N-body models (see Fig. 10a) is
 connected with the growing importance of statistical fluctuations
 with decreasing number of stars. For $N=250$ the number of stars in the core
 at the time of core bounce is extremely small (about 14 particles).
 Therefore stochastic processes connected with formation of binaries and
 their subsequent burning cannot be properly approximated by continuous
 formulae (Goodman 1984, 1987). Results obtained by Giersz \& Heggie (1993b)
 for stochastic binary formation and burning with an isotropic gaseous model
 suggest that the minimum is
 in this case much sharper than in the standard case, whereas the
 post-collapse expansion is nearly the same. On the basis of that result we
 conclude that statistical fluctuations are important and at
 least partially account for the sharper minimum.
 \mn
 It can be concluded that the core evolution for large particle numbers
 is excellently described by anisotropic gaseous models in pre- as
 well as in post-collapse with the standard phenomenological
 implementation of the binary energy fed into the core. Remaining
 problems are faced by the gaseous model in two respects: i) small
 particle numbers and three-body encounters: they tend to spread
 energy non-locally accompanied by a large generation of anisotropy, which
 cannot properly be taken into account by the present simple
 isotropic energy input
 formula for binaries in the gaseous model; ii) large particle numbers and
 two-body scatterings: in the regions outside the core of the system
 an appreciable difference occurs in the rate of evolution of the
 Lagrangian radii -- the gaseous model evolves too fast. Whether this
 is a result of approximations inherent in the gaseous model (e.g. local
 approximation for all collisional effects, or
 neglect of higher order moments of the velocity
 distribution) remains a question for
 future work. One could for example take into account non-local
 scattering in the context of a numerical solution of the
 orbit-averaged 2-D Fokker-Planck
 equation or a more detailed form of the
 velocity distribution by a higher order moment model.
 \vfill\eject
 \leftline{\bf ACKNOWLEDGMENTS}
 \pn
 Support by SERC grants No. GR/G04820 (M. Giersz) and No. GR/H56830
 (R. Spurzem) is gratefully acknowledged. R. Spurzem wants to thank
 Douglas C. Heggie for the invitation to come to Edinburgh and his
 continuous support which finally made this visit possible.
 Numerical computations were
 performed on two parallel supercomputers installed at the Edinburgh
 Parallel Computing Centre ; the $N=10000$
 $N$-body calculation is running on the HLRZ CRAY YMP-832 of KFA
 J"ulich, Germany. R.Sp. also wants to thank F.Flick F"orderungsstiftung
 (Germany) for financial support (travel grant) in 1991
 which helped initiating
 the project of the N-body calculations with S. Aarseth.
 We are indebted to J. Blair-Fish of Edinburgh
 Parallel Computer Centre for much help in launching N-body programs
 on the parallel computers.
 Both authors would like to thank D.C. Heggie for
 continuously supporting this project by many enlightening
 remarks and helpful discussions and for supplying the
 program used for most of the direct $N$-body calculations. We are
 also indebted to S. Aarseth, who supplied the NBODY5 program,
 which made possible the calculations for $N=2000$ and $N=10000$.
 \pn\vfill\eject
 \vskip18truept
 \leftline{\bf REFERENCES}
 \vskip8truept
 \parindent=0pt
 \pn
 \REF{\Text}{Aarseth S.J., 1985, in
 Brackbill J.U., Cohen B.I., eds, Multiple Time Scales.
 Academic Press, Orlando,
   p. 378}{}{}{}{}
\REF{\AA}{Aarseth S.J., H\'enon M., Wielen R.}{1974}{37}{183}{}
\REF{\ARAA}{Aarseth S.J., Lecar M.}{1975}{13}{1}{}
\REF{\MN}{Bettwieser E.}{1983}{203}{811}{}
\REF{\AA}{Bettwieser E., Spurzem R.}{1986}{161}{102}{}
\REF{\MN}{Bettwieser E., Sugimoto D.}{1984}{208}{439}{}
\REF{\MN}{Bettwieser E., Sugimoto D.}{1985}{212}{189}{}
\REF{\ApJ}{Cohn, H., Kulsrud, R.M.}{1978}{226}{1087}{}
\REF{\ApJ}{Cohn, H.}{1979}{234}{1036}{}
\REF{\ApJ}{Cohn, H.}{1980}{242}{765}{}
\REF{\Text}{Cohn H., 1985, in Goodman J., Hut P., eds,
 Proc. IAU Symp. 113, Dynamics of Star Clusters.
 Reidel, Dordrecht,
 p. 161}{}{}{}{}
\REF{\ApJ}{Cohn H., Hut P., Wise M.}{1989}{342}{814}{}
\REF{\Text}{Giersz M., Heggie D.C., 1993a, MNRAS, submitted}{}{}{}{}
\REF{\Text}{Giersz M., Heggie D.C., 1993b, MNRAS, to be submitted}{}{}{}{}
\REF{\ApJ}{Goodman J.}{1984}{280}{298}{}
\REF{\ApJ}{Goodman J.}{1987}{313}{576}{}
\REF{\Text}{Goodman J., Heggie D.C., Hut P., 1993, ApJ, submitted}{}{}{}{}
\REF{\Text}{Heggie D.C., 1973, in  Tapley B.D., Szebehely V., eds,
Recent Advances in Dynamical Astronomy. Reidel, Dordrecht}{}{}{}{}
\REF{\MN}{Heggie D.C.}{1984}{206}{179}{}
\REF{\MN}{Heggie D.C., Stephenson D.}{1988}{230}{223}{}
\REF{\MN}{Heggie D.C., Ramamani N.}{1989}{237}{757}{}
\REF{\Text}{Janes K., ed, 1991, The Formation and Evolution of
Star Clusters. A.S.P., San Francisco}{}{}{}{}
\REF{\Nat}{Hut P., Makino J., McMillan S.L.W.}{1988}{336}{31}{}
\REF{\MN}{Larson R.B.}{1970}{147}{323}{}
\REF{\MN}{Louis P.D.}{1990}{244}{478}{}
\REF{\MN}{Louis P.D., Spurzem R.}{1991}{244}{478}{(LS)}
\REF{\AJ}{Lupton R.H., Gunn J.E.}{1987}{93}{1106}{}
\REF{\AJ}{Lupton R.H., Gunn J.E., Griffin R.F.}{1987}{93}{1114}{}
\REF{\MN}{Lynden-Bell D., Eggleton P.P.}{1980}{191}{483}{}
\REF{\ApJ}{Miller R.H.}{1964}{140}{250}{}
\REF{\Text}{Press W.H., Flannery B.P., Teukolsky S.A, Vetterling W.T.,
 1986, Numerical Recipes. Cambridge Univ. Press, Cambridge}{}{}{}{}{}
\REF{\Text}{Rosenbluth M.N., McDonald W.M., Judd D.L., 1957,
  Phys. Rev., 107, 1}{}{}{}{}
\REF{\Text}{Saslaw W.C., 1985, Gravitational Physics of Stellar and
Galactic Systems. Cambridge Univ. Press, Cambridge}{}{}{}{}
\REF{\Text}{Spitzer L.,  1987, Dynamical Evolution of Globular
 Clusters, Princeton Univ. Press, Princeton}{}{}{}{}{}
\REF{\ApJ}{Spitzer L., Mathieu R.D.}{1980}{241}{618}{}
\REF{\Text}{Spurzem R., Aarseth S.J., 1993, in preparation}{}{}{}{}
\REF{\Text}{Spurzem R., Louis P.D., 1993, in preparation}{}{}{}{}
\REF{\Acta}{Stod\'o\l kiewicz J.S.}{1982}{32}{63}{}
{}{}{}{}{}
\vfill\break
\bn
\centerline{\bf APPENDIX}
\bn
\centerline{\bf Derivation of the collisional decay of anisotropy}
\bn
 According to Larson (1970) we approximate the real local
 (r-dependent) velocity distribution function $f$ as a function
 of the modulus of
 a star's random velocity vector $v$
 and the cosine of its angle with the radial direction $\mu$,
 $$ f = \sum_{j=0}^{\infty} A_j( {v^\prime} ) P_j (\mu^\prime)
    \eqno(A1) $$
 where
 $P_j(\mu ) $ denotes the Legendre polynomials of order $j$.
 We are going to apply the Fokker-Planck equation
 in $v,\mu $ coordinates, which was computed by Rosenbluth, McDonald \& Judd
 (1957, RMB); therefore we need to know the Rosenbluth potentials

  $$ h(\vvec) = 2m \int f({\bf v}^{\prime})
    \vert\vvec - {\bf v}^{\prime}\vert^{-1} d^3{\bf v}^{\prime}
    \eqno(A2)      $$

 $$ g(\vvec) = m \int f({\bf v}^{\prime})
    \vert\vvec - {\bf v}^{\prime}\vert d^3{\bf v}^{\prime}
   \eqno(A3)      $$
 for the distribution function of Eq. (A1); they are
 $$ h(v,\mu) = 8\pi m\cdot \sum_{l=0}^{\infty}
        {P_l(\mu)\over 2l+1 }
 \intl_0^{\infty} \vquot A_l( v^\prime ) v^{\prime 2} dv^\prime
 \eqno(A4) $$

  $$ \eqalign{ g(v,\mu) =
      & 4\pi m \cdot \sum_{l=0}^{\infty}{1\over 2l+1}
  \cdot\cr \Biggl\{ &P_l(\mu) \intl_0^{\infty}
   A_l( v^\prime )(v^2+v^{\prime 2})v^{\prime 2}
   \vquot dv^\prime \cr
    -&P_{l-1}(\mu){2l\over 2l-1}\intl_0^{\infty}
   A_{l-1}( v^\prime )vv^{\prime 3}
   \vquot dv^\prime \cr
    -&P_{l+1}(\mu){2l+2\over 2l+3} \intl_0^{\infty}
   A_{l+1}( v^\prime )vv^{\prime 3}
   \vquot dv^\prime \Biggr\} \cr} \eqno(A5) $$
 with $v_< = \min(v,v^\prime)$, $v_> = \max(v,v^\prime)$.
 \mn
 Since we are interested mainly in distribution functions
 with anisotropy in the second order moments the series in
 Eq. (A1) shall be truncated now at that order and we use
 in accord with Larson's (1970) definitions
 $$\eqalign{A_0& = g(v) \cr
            A_1& = 0 \cr
            A_2&= c_2 {v^2\over\sigma^2} g(v) \cr} \eqno(A6) $$
 with
   $$  c_2  = {\sr^2-\s_t^2\over 3\sigma^2 }\ .   \eqno(A7) $$
 $$ g(v) = {1\over\sqrt{2\pi}^3\sigma^3}\exp\Bigl(-{v^2\over 2\sigma^2}\Bigr)
               \eqno(A8) $$
 is the isotropic Maxwell-Boltzmann distribution.
 These definitions just ensure that the quantities $\sr^2$, $\s_t^2$,
 and $3\sr^2(v_r-u)$ are recovered by determining the second
 and third order moments of $f$;
 fourth order moments have been assumed to take the
 same value as in the case of a Maxwell-Boltzmann distribution
 function; in Larson's notation this means that we have set
 $\xi =0$ (see his Eq. 2).
 \mn
 Now the Rosenbluth potentials can be further evaluated in terms
 of the functions
 $$I_n = \intl_0^v
            v^{\prime n} g( v^\prime ) dv^\prime \eqno(A9) $$

 $$K_n = \intl_v^{\infty}
            v^{\prime n} g(v^\prime) dv^\prime \ . \eqno(A10) $$
 As a final result we get
 $$
  h(v,\mu) = 8\pi m
 \ \Biggl\{ \ \Bigl[{1\over v}I_2 + K_1\Bigr] +
  {1\over 5} P_2(\mu) {c_2\over \sigma^2}
   \ \Bigl[{1\over v^3}I_6 + v^2 K_1\Bigr]
  \ \Biggr\} \eqno(A11) $$
 $$\eqalign{ g(v,\mu) = 4\pi m\Biggl\{\
   &\Bigl[vI_2+{1\over 3v}I_4 +{v^2\over 3} K_1 + K_3 \Bigr] \cr
  +{1\over 5}{c_2\over\sigma^2}P_2(\mu)\
   &\Bigl[-{1\over 3v}I_6+{1\over 7v^3}I_8
          +{v^4\over 7}K_1-{v^2\over 3}K_3\Bigr]
  \ \Biggr\} \cr
   }  \eqno(A12) $$
 With these potentials the Fokker-Planck equation of RMB is
 determined; let the right hand side of their Eq. (31) be denoted with
 FP; we then evaluate
 $$ \eqalign{   Dp_r =
  \dedet{p_r}{c} &= 2\pi\rho\intl_0^\infty v^2 dv \intl_{-1}^{+1} d\mu
    v^2\mu^2 {\,\rm FP}\cr
      Dp_t =
  \dedet{p_t}{c} &= 2\pi\rho\intl_0^\infty v^2 dv \intl_{-1}^{+1} d\mu v^2
               {1\over 2} (1-\mu^2) {\,\rm FP} \cr} \eqno(A13) $$
 It turns finally out that
 $$ Dp_r - Dp_t = -{p_r - p_t\over T} \Bigl( {9\over 10} -
         {9\over 140}{p_r-p_t\over p }
        \Bigr)   ,  \eqno(A14) $$
 with the average pressure $p=(p_r+2p_t)/3 = $ $\rho\sigma^2 $.
 Since the second term in the above equation is small for
 reasonable values of the anisotropy we finally
 get
 $$ T_{\rm A} = -{p_r-p_t\over Dp_r - Dp_t} =  {10\over 9} T . \eqno(A15) $$
 \vfill\eject

 {\bf FIGURE CAPTIONS}
 \medskip

 {\bf Fig.1} Evolution of Lagrangian radii containing the stated percentage
 of the total mass for an isotropic (IGM) and anisotropic
 (AGM) gaseous model for a $N=1000$ star cluster

 \smallskip

 {\bf Fig.2a} Radial profile of the logarithmic density power law index
 $\alpha$ for a number of gaseous models with different
 evolutionary time. The dash-dotted line represents the self-similar
 anisotropic pre-collapse model of LS

 \smallskip

 {\bf fig.2b} Radial profile of the anisotropy $A$
 for a number of gaseous models with different evolutionary time. The
 dash-dotted line represents the self-similar anisotropic pre-collapse
 model of LS

 \smallskip

 {\bf Fig.3a} Comparison between the density power law index $\alpha$ as a
 function of $\lambda_A \lambda$ for  anisotropic gaseous models (AGM) and
 self-similar anisotropic pre-collapse models
 of LS (SIM). The values of $\alpha$ for the isotropic gaseous model of
 Lynden-Bell  and Eggleton (LBE) and the isotropic Fokker-Planck model of
 Cohn (FP) are marked

 \smallskip

 {\bf Fig.3b} Comparison between the anisotropy $A$ as a
 function of $\lambda_A \lambda$ for  anisotropic gaseous models (AGM) and
 self-similar anisotropic pre-collapse models  of LS (SIM)

 \smallskip

 {\bf Fig.4} Amount of shift necessary to adjust the averaged
 $N=1000$ $N$-body model to the anisotropic
 gaseous model for $1$\% Lagrangian radius

 \smallskip

 {\bf Fig.5} Evolution of the 1\% Lagrangian radius in the averaged
 $N=1000$  $N$-body model in comparison to the anisotropic gaseous model
 for different initial times $t_{b0}$ for switching
 on the binary energy generation.  The $cc$ indicates a pure core
 collapse gaseous model

 \smallskip

 {\bf Fig.6a} Evolution of the 1\% Lagrangian radius in the averaged
 $N=1000$ $N$-body model in comparison to the anisotropic gaseous model
 for different strength of the binary energy generation parameter
 $C_b$. The $cc$ indicates a pure core collapse gaseous model

 \smallskip

 {\bf Fig.6b} Time variation of $x$ (Def. see text) for three
 particle   numbers indicated at the curves, in the averaged
 $N$-body models; all times were normalized such
 that the core-collapse time is the same for all $N$ and the
 strongly fluctuating curves were smoothed.

 \smallskip

 {\bf Fig.7} Evolution of the $1$\% Lagrangian radius in the
 averaged $N=1000$, $N=500$, $N=250$  $N$-body models
 in comparison to the anisotropic gaseous models
 $(C_b = 90,\ t_{b0} = 230)$,
 $(C_b = 70,\ t_{b0} = 130)$,
 $(C_b = 55,\ t_{b0} = 50)$, respectively

 \smallskip

 {\bf Fig.8a} Evolution of the $1$\%, $2$\%, $5$\%, $10$\%
 Lagrangian radii in the averaged
 $N=1000$ $N$-body model
 in comparison to the anisotropic gaseous model

 \smallskip

 {\bf Fig.8b} Evolution of the $20$\%, $50$\%, $75$\% Lagrangian radii
 in the averaged $N=1000$ $N$-body model
 in comparison to the anisotropic gaseous model

 \smallskip

 {\bf Fig.9a} The same as in Fig. 8a but for $N=500$

 \smallskip

 {\bf Fig.9b} The same as in Fig. 8b but for $N=500$

 \smallskip

 {\bf Fig.10a} The same as in Fig. 8a but for $N=250$

 \smallskip

 {\bf Fig.10b} The same as in Fig. 8b but for $N=250$. The
 $75$\% Lagrangian radius is not available for this model.

 \smallskip

 {\bf Fig.11} Evolution of the $90$\% Lagrangian radius in the
 averaged $N=2000$ $N$-body model in comparison to the anisotropic
 gaseous model

 \smallskip

 {\bf Fig.12} Evolution of the anisotropy between Lagrangian radii of
 $10$\% to $20$\% ($A_5$),
 $20$\% to $50$\% ($A_6$), and $50$\% to $100$\%
 ($A_7$) in the averaged $N=1000$
 $N$-body model  in comparison to the anisotropic
 gaseous models for different
 parameters of collisional anisotropy decay, $\lambda_A$

 \smallskip

 {\bf Fig.13} The same as in Fig. 12 but for $N=500$

 \smallskip

 {\bf Fig.14} The same as in Fig. 12 but for $N=250$

 \smallskip

 {\bf Fig.15} Evolution of the anisotropy between Lagrangian radii of
 $40$\% to $50$\%, $50$\% to $75$\% and $75$\% to $90$\% in the
 averaged $N=2000$ $N$-body model in comparison to the anisotropic gaseous
 model  without binaries; note the different definition of Lagrangian
 shells here.

 \smallskip

 {\bf Fig.16} The same as in Fig. 12 but for $N=10000$ and $\lambda_A=0.1$
 and no binaries

 \smallskip

 {\bf Fig.17} Evolution of the anisotropy between Lagrangian radii of
 $10$\% to  $20$\%, $20$\% to $50$\% and $50$\% to $100$\% in the
 averaged $N=1000$ $N$-body model  in comparison to the anisotropic gaseous
 models for different $\beta$ (see text)

 \smallskip

 {\bf Fig.18} The same as in Fig. 17 but for $N=500$

 \smallskip

 {\bf Fig.19} Evolution of the anisotropy between Lagrangian radii of
 $50$\% to $100$\% in the
 averaged $N=1000$ $N$-body models with different
 smoothing parameters $\epsilon$  in comparison
 to the anisotropic gaseous  model

 \bye